\documentclass[12pt]{article}
\pdfoutput=1
\usepackage{jcappub}

\newcommand{\be}{\begin{equation}}
\newcommand{\ee}{\end{equation}}
\begin{document}
\subheader{SU-ITP-11/45, SLAC-PUB-14555}
\title{Analytic Coleman-De Luccia Geometries}
\author[a,b]{Xi Dong,}
\author[a]{Daniel Harlow}
\affiliation[a]{Stanford Institute for Theoretical Physics,\\
Department of Physics, Stanford University,\\
Stanford CA 94305 U.S.A.}
\affiliation[b]{Theory Group, SLAC National Accelerator Laboratory,\\
2575 Sand Hill Road, Menlo Park, CA 94025 U.S.A.}
\emailAdd{xidong@stanford.edu}
\emailAdd{dharlow@stanford.edu}

\abstract{We present the necessary and sufficient conditions for a Euclidean scale factor to be a solution of the Coleman-De Luccia equations for some analytic potential $V(\phi)$, with a Lorentzian continuation describing the growth of a bubble of lower-energy vacuum surrounded by higher-energy vacuum.  We then give a set of explicit examples that satisfy the conditions and thus are closed-form analytic examples of Coleman-De Luccia geometries.}

\maketitle

\section{Introduction} The Coleman-De Luccia (CDL) geometry \cite{CDL} is essential to the study of eternal inflation (see \cite {Garriga:1997ef,Guth:2007ng} and the references therein) and the string theory landscape \cite{Bousso:2000xa,Kachru:2003aw,Susskind:2003kw}.  Most discussions of this geometry take place in the ``thin-wall'' limit, where the geometries inside and outside of the bubble are pieces of de Sitter space, Minkowski space, or Anti de Sitter space, sewn together nonanalytically at the domain wall.  This picture is sufficient for many qualitative questions, but is troublesome when applied to calculations of correlation functions in the CDL background \cite{fssy,Dong:2011uf}.  This often involves various analytic continuations from Euclidean to Lorentzian signature, and also from one part to another of the geometry, and pathologies can appear when the geometry is not analytic.  Our goal in this paper is to give some reasonably simple analytic expressions for ``thick-wall'' CDL bubbles which mediate various types of decays.  There is a conservation of trouble here, however, in that we will not be able to give closed form expressions for the scalar potentials which give rise to these geometries.  Rather we will work out a set of general constraints that the metric needs to obey in order for it to come from \textit{some} scalar field theory with a potential, and then give examples of geometries which satisfy all the constraints.  Since it is the potential that usually comes out of ``top-down'' constructions this may not seem eminently useful, but we consider our approach to be more appropriate for ``bottom-up'' investigations of bulk physics in the CDL background.  Similar analyses have been applied to other less-restricted cases such as thick domain walls \cite{DeWolfe:1999cp,Gremm:1999pj,Csaki:2000fc} and FRW cosmologies \cite{Ellis:1990wsa}, both of which are contained in our analytic CDL geometries.

\section{General Properties of the Coleman-De Luccia Geometry}
\subsection{Euclidean Preliminaries}
The Euclidean CDL geometry is a solution of the equations of motion for the scalar/gravity system with Euclidean action
\be
S=-\frac{1}{16\pi G}\int d^d x\sqrt{g}R+\int d^d x\sqrt{g}\left[\frac{1}{2}g^{\mu\nu}\partial_\mu \phi \partial_\nu \phi+V(\phi)\right].
\ee
The solution has $SO(d)$ symmetry, so we can write the metric as 
\be
ds^2=d\xi^2+a(\xi)^2 \left(d\theta^2+\sin^2 \theta d\Omega_{d-2}^2\right).
\ee
The equations of motion for solutions with this symmetry are
\begin{gather}\nonumber
\phi''+(d-1)\frac{a'}{a}\phi'-V'(\phi)=0\\
\left(\frac{a'}{a}\right)^2=\frac{1}{a^2}+\frac{16\pi G}{(d-1)(d-2)}\left(\frac{1}{2}\phi'^2-V(\phi)\right).\label{eom}
\end{gather}

If $V$ has a local extremum at $\phi=\phi_{min}$ then there is a simple solution.  When $V(\phi_{min})=0$ we have flat space:
\be\label{eflata}
a(\xi)=\xi.
\ee
When $V(\phi_{min})=\rho_{min}>0$, we have the sphere:
\begin{align}\nonumber
a(\xi)&=\ell_{ds}\sin(\xi/\ell_{ds})\\
\ell_{ds}^{-2}&=\frac{16\pi G \rho_{min}}{(d-2)(d-1)}.\label{edsa}
\end{align}
When $V(\phi_{min})=\rho_{min}<0$, we have hyperbolic space:
\begin{align}\nonumber
a(\xi)&=\ell_{ads}\sinh(\xi/\ell_{ads})\\
\ell_{ads}^{-2}&=-\frac{16\pi G \rho_{min}}{(d-2)(d-1)}.\label{eadsa}
\end{align}

More interesting solutions will interpolate smoothly between different minima of the potential; these have the interpretation of causing bubble nucleation.\footnote{The Euclidean solution does not actually quite reach the minimum; the more precise boundary conditions are stated momentarily.  Also for vacua which have $V<0$ it is possible for a maximum to be stable or metastable; we include this case below in our definition of a CDL geometry.}  The geometries describing the decay of Minkowski space and AdS are noncompact and can be chosen to have $\xi\in [0,\infty)$, while the geometry describing the decay of dS space is compact and can be chosen to have $\xi \in [0,\xi_c]$.  In all cases for the solution to be smooth at $\xi=0$ we need
\begin{align}\nonumber
\phi(\xi)&=\phi_0+\mathcal{O}(\xi^2)\\
a(\xi)&=\xi+\mathcal{O}(\xi^3).\label{bcat0}
\end{align}
For the noncompact cases, as $\xi\to\infty$ we want $\phi$ to approach its value in the false vacuum and $a$ to approach (\ref{eflata}) or (\ref{eadsa}).  When the false vacuum is dS then as $\xi \to \xi_c$ smoothness requires
\begin{align}\nonumber
\phi(\xi)&=\phi_c+\mathcal{O}((\xi_c-\xi)^2)\\
a(\xi)&=(\xi_c-\xi)+\mathcal{O}((\xi_c-\xi)^3).\label{bcat1}
\end{align}

\subsection{Lorentzian Continuation}
\begin{figure}[!htbp]
\includegraphics[scale=.9]{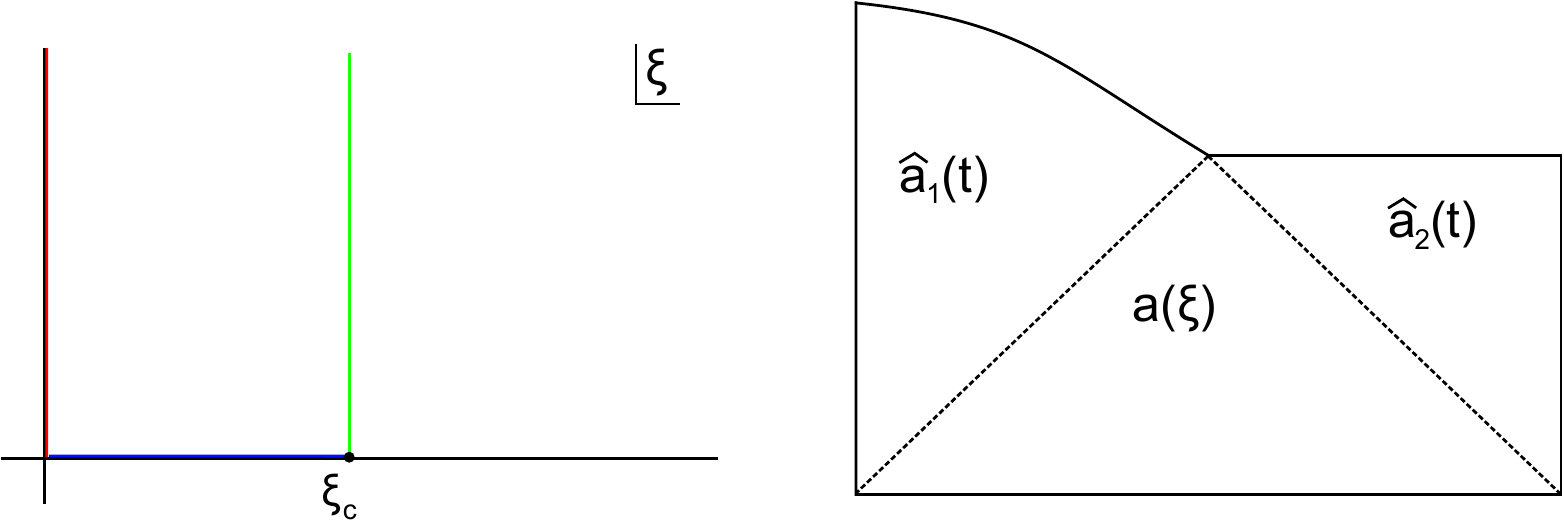}
\caption{On the left we have the three regions of interest in the $\xi$ plane for the Lorentzian continuation of a compact CDL geometry.  The red line gives $\hat{a}_1(t)$, the blue line gives $a(\xi)$, and the green line gives $\hat{a}_2(t)$.  On the right we show the regions of the CDL Penrose diagram that are described by these different continuations.  If the geometry is noncompact then the blue line extends to infinity and the $\hat{a}_2$ region doesn't exist.}
\label{fig:continuation}
\end{figure}

To find the Lorentzian geometry describing the aftermath of bubble nucleation we analytically continue the Euclidean solution of the previous subsection.  To get inside the bubble we define
\begin{align}\nonumber
\xi&=it\\\nonumber
\theta&=i \rho\\\label{con1}
\hat{a}_1(t)&=-i a(it),
\end{align}
which gives an open FRW cosmology
\be\label{frw}
ds^2=-dt^2+\hat{a}_1(t)^2 \left(d\rho^2+\sinh^2 \rho d\Omega_{d-2}^2\right).
\ee
To get the Lorentzian geometry outside of the bubble we continue 
\be \label{con2dw}
\theta=\frac{\pi}{2}+i\omega,
\ee
which gives a ``warped de Sitter'' geometry
\be\label{warpedds}
ds^2=d\xi^2+a(\xi)^2\left[-d\omega^2+\cosh^2 \omega d\Omega_{d-2}^2\right].
\ee
When the false vacuum is dS there is an additional region outside of the bubble which is up near the future boundary of the false vacuum, which we reach by 
\begin{align}\nonumber
\xi&=\xi_c+it\\\nonumber
\theta&=i \rho\\\label{con2}
\hat{a}_2(t)&=i a(\xi_c+it).
\end{align}
Both $\hat{a}_1$ and $\hat{a}_2$ obey the Lorentzian FRW equations of motion
\begin{gather}\nonumber
\ddot{\phi}+(d-1)\frac{\dot{\hat{a}}}{\hat{a}}\dot{\phi}+V'(\phi)=0\\\label{leom}
\left(\frac{\dot{\hat{a}}}{\hat{a}}\right)^2=\frac{1}{\hat{a}^2}+\frac{16\pi G}{(d-1)(d-2)}\left(\frac{1}{2}\dot{\phi}^2+V(\phi)\right).
\end{gather}
The metric and scalar produced by these continuations are guaranteed to be real because they obey equations of motion and boundary conditions that are real.  In particular the simple Euclidean solutions with constant $\phi$ become various patches of Minkowski, de Sitter, or Anti de Sitter space.  The late time behaviour inside of the bubble depends on the nature of the ``true'' minimum.  If the minimum is infinitely far away in field space then there are many possibilities, but if the minimum is at some finite $\phi$ there are only three.  These are Minkowski space\footnote{We use ``$\to$'' here instead of ``$\sim$'' to indicate $a(t)=t(1+o(t^0))$, i.e.\ unlike the dS and AdS cases the prefactor must go to one.}
\be \label{lflata}
\hat{a}_1(t)\to t,
\ee
de Sitter
\be\label{ldsa}
\hat{a}_1(t)\sim e^{t/\ell_1},
\ee
and Anti de Sitter\footnote{In this case late time does not make so much sense, since we expect the geometry to crunch in time of order $\ell_1$ and there isn't really a good asymptotic limit.}
\be\label{ladsa}
\hat{a}_1(t)\sim \sin (t/\ell_1).
\ee
If the false vacuum has positive energy and is at finite field value then we also expect
\be\label{ldsa2}
\hat{a}_2(t)\sim e^{t/\ell_2}.
\ee
The full analytic continuation is illustrated in Figure \ref{fig:continuation}.

\subsection{Constraints and a Definition}
For a given potential the boundary conditions are sufficient to determine a unique solution of the equations of motion (\ref{eom}).  What we will do in this section is to identify the constraints that a real and positive function $a(\xi)$ must obey in addition to the boundary conditions to ensure that a potential exists which has this $a(\xi)$ (and some $\phi(\xi)$) as a solution, and also that its analytic continuation describes a Lorentzian bubble geometry of true vacuum surrounded by false vacuum.  

We begin by writing expressions for $\phi$ and $V$ in terms of $a$:
\begin{align}\nonumber
\frac{8\pi G}{d-2}\phi'^2&=\left(\frac{a'}{a}\right)^2-\frac{1}{a^2}-\frac{a''}{a}\\\label{phiVeq}
\frac{16\pi G}{(d-2)^2}V(\phi)&=\frac{1}{a^2}-\left(\frac{a'}{a}\right)^2-\frac{1}{d-2}\frac{a''}{a}.
\end{align}
The first of these make it clear that throughout the physical range of $\xi$ we must have
\be\label{enec}
a'^2-a a''-1\geq0.
\ee
In fact if this inequality is satisfied then we can integrate the first equation in \eqref{phiVeq} to find $\phi(\xi)$, which we can then invert and insert into the second equation to find $V(\phi)$.\footnote{If there are places where the inequality is saturated and $\phi$ comes to a rest, this inversion is slightly more subtle.  This happens for example inside the bubble if the field oscillates about the true minimum before settling down, as in reheating.  This subtlety does not affect our ability to find the potential since the scalar traverses all relevant parts of the potential at least once.}  This inequality may appear unusual, but we show in Appendix \ref{necapp} that it is equivalent to the null energy condition in the region produced by the continuation (\ref{con2dw}). 

Additional constraints come from the Lorentzian continuation.  We want $\hat{a}_1(t)$ to be real and positive for all $t>0$, and also to obey (\ref{lflata}), (\ref{ldsa}), or (\ref{ladsa}), with the caveat that in the last case \eqref{ladsa} $\hat a_1(t)$ need only be positive before the crunch.  If the false vacuum is dS then we want $\hat{a}_2(t)$ to be real and positive for all $t>0$ and to obey (\ref{ldsa2}).  By continuing (\ref{phiVeq}) we see that to reconstruct the scalar field and potential we need both $\hat{a}_1$ and $\hat{a}_2$ to obey
\be
\label{lnec}
\dot{\hat{a}}^2-\hat{a} \ddot{\hat{a}}-1\geq0.
\ee
This inequality is again equivalent to the null energy condition in these two regions.  We claim that these are all of the constraints that a proposed $a(\xi)$ needs to obey to be considered a CDL geometry.

The condition that $\hat{a}_1(t)$ is real can be rewritten in a nice way by observing that in a neighborhood around $\xi=0$ we can see from the Taylor expansion for $a(\xi)$ that it is equivalent to
\be\label{refl1}
a(-\xi)=-a(\xi).
\ee
By analytic continuation this equation must hold in any simply-connected region containing $\xi=0$ in which $a(\xi)$ is analytic.  Conversely, \eqref{refl1} (together with the reality and positivity of $a(\xi)$ on $(0,\xi_c)$) implies that $\hat a_1(t)$ is real and positive for small positive $t$, and by analytic continuation must be so for all $t>0$ unless we encounter a zero or singularity.  Similarly the condition that $\hat{a}_2(t)$ is real is equivalent to
\be\label{refl2}
a(\xi_c-\xi)=-a(\xi_c+\xi).
\ee

A final condition we would like to have is that the vacua involved are at least metastable and not unstable.  If the geometry is compact then the late time behaviour (\ref{lflata}-\ref{ldsa2}) of $\hat{a}_{1,2}$ ensures that the scalar field is rolling down to a minimum in both asymptotic regions.  But if the geometry is noncompact then the false vacuum is reached already in the Euclidean geometry as $\xi\to\infty$, and there is no $\hat{a}_2$.  So if we do not impose an additional condition, then the constraints we have stated so far allow situations where there is an unstable Minkowski or AdS maximum that the field rolls down from, which we feel does not deserve the name of a CDL geometry since it is not a tunnelling process.  If the false vacuum is Minkowski we thus need to demand that $V''(\phi(\xi))>0$ as $\xi\to\infty$.  The simplest way to achieve this is to demand that the potential is decreasing as $\xi\to\infty$, which from \eqref{phiVeq} means that
\be
\label{ncVmin}
\left(\frac{1}{a^2}-\left(\frac{a'}{a}\right)^2-\frac{1}{d-2}\frac{a''}{a}\right)'<0\qquad \mathrm{as} \,\,\,\,\, \xi\to\infty.
\ee
Note that unlike our other restrictions, this one depends on the dimension $d$.  If the false vacuum is AdS, then a maximum should be allowed if its negative mass-squared obeys the Breitenlohner-Freedman bound \cite{Breitenlohner:1982jf} 
\be\label{bfbound}
V''(\phi(\xi))>-\frac{(d-1)^2}{4(\ell_2)^2}\qquad \mathrm{as}\,\,\,\,\,\xi\to\infty.
\ee  
The potential in this inequality is written in terms of $a$ by using (\ref{phiVeq}).  Examples of potentials with metastable maxima of this type were given in \cite{Hertog:2004rz,Hertog:2005hu}.

We can now gather the results of this section into a definition; a function $a(\xi)$ is a ``CDL Geometry'' if:
\begin{itemize}
\item[(a)] It is real and positive on a real interval $\xi\in (0,\xi_c)$, possibly with $\xi_c\to \infty$.  
\item[(b)] Near $\xi=0$ it obeys (\ref{bcat0}), and if $\xi_c$ is finite then it obeys (\ref{bcat1}).  If $\xi_c$ is infinite then as $\xi\to \infty$ either $a(\xi)\sim e^{\xi/\ell_2}$ (``tunnelling from AdS'') or $a(\xi)\to\xi$ (``tunnelling from Minkowski'').
\item[(c)] It is analytic in a simply connected region $\mathcal{D}$ containing the real interval $[0,\xi_c]$, the positive imaginary axis,\footnote{For tunnelling to (crunching) AdS we only include the open interval between the origin and the first zero of $a$ on the positive imaginary axis. This also applies to constraints (d) and (e).} and if $\xi_c$ is finite the ray $\mathcal{R}$ defined by $\xi=\xi_c+it$ with $t>0$.
\item[(d)] There are no zeros of $a$ on the positive imaginary axis or on $\mathcal{R}$ if $\xi_c$ is finite, and throughout $\mathcal{D}$ we have $a(-\xi)=-a(\xi)$.  If $\xi_c$ is finite we also have $a(\xi_c-\xi)=-a(\xi_c+\xi)$ and the dS asymptotic (\ref{ldsa2}).  
\item[(e)] On the real interval $[0,\xi_c]$ we have the null energy condition \eqref{enec}. On the positive imaginary axis and on $\mathcal{R}$ if $\xi_c$ is finite, we have the null energy condition (\ref{lnec}).
\item[(f)] If $\xi_c$ is infinite then for ``tunnelling from Minkowski'' the inequality \eqref{ncVmin} is satisfied, while for ``tunnelling from AdS'' we have \eqref{bfbound}.
\end{itemize}
In this definition we assumed that the false vacuum was at a finite point in field space.  If we wish to restrict to cases where the true vacuum is also at a finite point in field space then we can introduce an additional requirement:
\begin{itemize}
\item[(g)] For large purely imaginary $\xi$ we have the asymptotic geometry (\ref{lflata}), (\ref{ldsa}), or (\ref{ladsa}). 
\end{itemize}

\subsection{Compact Coleman-De Luccia Geometries}\label{comp}
Before presenting our examples of CDL geometries, we will make some special observations about the compact case.  We first note that the reality conditions (\ref{refl1}) and (\ref{refl2}) allow continuation of $a(\xi)$ to a neighborhood of the full real axis and together imply the periodicity
\be
a(\xi+2\xi_c)=a(\xi).
\ee
Thus for the compact case we have Fourier Analysis at our disposal.  We will henceforth choose units where $\xi_c=\pi$, after which we see that we may write
\be
a(\xi)=\sum_{n=1}^\infty c_n \sin(n\xi).
\ee
There are no cosines because the function is odd.  This form is not the most useful however because the boundary conditions (\ref{bcat0}) and (\ref{bcat1}) imply nontrivial constraints on the $c_n$'s.  We can reorganize the series using trigonometric identities to take the form
\be\label{sc}
a(\xi)=\sin(\xi)\left[1+f(\sin \xi)+\cos(\xi)g(\sin\xi)\right],
\ee
where $f(\cdot)$ and $g(\cdot)$ are even functions which go to zero as their argument goes to zero, and which are analytic in a region containing the real interval $(0,1]$ and also the pure imaginary axis.  This form is completely general; any compact CDL geometry must have it.  It is convenient because it automatically incorporates the reality conditions and boundary conditions at $0$ and $\pi$, so the only remaining things to check are that the scale factor is nonvanishing, the null energy condition is satisfied, and that at late times outside the bubble we have (\ref{ldsa2}).  The cosine term has a simple interpretation in that it breaks the symmetry between the two minima.  In particular, the analytic continuations \eqref{con1} and \eqref{con2} give
\begin{equation}\label{shch}
\hat a_{1,2}(t)=\sinh(t)\left[1+f(i\sinh t)\pm\cosh(t)g(i\sinh t)\right],
\end{equation}
which are real because $f$ and $g$ are even functions. Here $\hat a_1$ (or $\hat a_2$) takes the plus (or minus) sign.  Depending on the desired nature of the true vacuum we might also like to impose (\ref{lflata}), (\ref{ldsa}), or (\ref{ladsa}).

\section{Examples}
In this section, we first derive some general results that illustrate the difficulty of constructing geometries that satisfy our definition.  In the following subsections we then give a series of explicit examples.  

In writing down functions $a(\xi)$ that obey our constraints (a)--(g), the biggest challenge is the null energy conditions (\ref{enec}) and (\ref{lnec}).  We first consider the case where $a(\xi)$ (or $\hat{a}(t)$) is linear with unit coefficient both at $\xi=0$ and $\xi\to\infty$.  This is appropriate for tunnelling to or from Minkowski space.  We parametrize this as
\be
a(\xi)=\xi(1+\delta(\xi)).
\ee
Here $\delta(\xi)$ must go to zero both at $\xi=0$ and $\xi=\infty$, and we must have $\delta>-1$ to avoid collapse.  If there is a point where $-1<\delta<0$, then by continuity $\delta$ must have a minimum $\xi_{min}$ with $-1<\delta(\xi_{min})<0$.  The null energy condition however takes the form
\be\label{deltanec}
\delta(2+\delta)+\xi^2\delta'^2-\delta''\xi^2(1+\delta)\ge0,
\ee
and it is easy to check that a local minimum with $-1<\delta<0$ necessarily violates it.  Thus we must have 
\be\label{deltabound}
\delta\geq0.
\ee
We can use this observation to constrain the behaviour of $a$ near $\xi=0$.  We can expand $\delta(\xi)=A\xi^n+\mathcal{O}(\xi^{n+2})$, with $n$ some even integer greater than 1.  Inserting this into \eqref{deltanec} we find that if $n>2$ then we must have $A<0$, which is not allowed because then near $\xi=0$ we would have $\delta<0$.  So $\delta(\xi)$ must start out like $A\xi^2$ with $A>0$.  We can also rule out the possibility that $\delta$ is a rational function.  If $\delta$ were rational then at large $\xi$ it would scale like $\xi^{-n}$ for some positive integer $n$.  From \eqref{deltanec}, we find the restriction $n(n+1)\leq2$.  The only possibility is $n=1$, but if $\delta$ is rational then this is impossible since (\ref{refl1}) requires $\delta$ to be an even function of $\xi$.  In our examples below we overcome this by including radicals.

Another constraint of this type is that when the false vacuum is de Sitter, its radius must be greater than 1 in the units where $\xi_c=\pi$:
\be
\ell_2>\frac{\xi_c}{\pi}.
\ee 
In the thin-wall limit, this is the statement that if we draw the Euclidean CDL instanton as a piece of a sphere glued to a piece of flat space, the hyperbolic plane, or a larger sphere, the radius of the sphere corresponding to the metastable dS vacuum is always larger than the ``size'' $\xi_c$ of the instanton divided by $\pi$.\footnote{One can also think of this as a bound on the proper length $\xi_c$ of the warped de Sitter region \eqref{warpedds}.}  We show this is generally true in Appendix \ref{dsrad}.  Note that when the ``true'' vacuum is also dS this also implies that $\ell_1>1$ since $\ell_1>\ell_2$ by definition.  This constraint implies that the even functions $f(\cdot)$ and $g(\cdot)$ in equation (\ref{sc}) cannot both be rational, as this would lead to scaling (\ref{ldsa2}) with $\ell_2^{-1}=(n+1)$ and $n\in \mathbb{Z}$ (we are now setting $\xi_c=\pi$).  This is clearly impossible if $\ell_2>1$.  So again in the compact case we will include radicals to avoid this problem.

\subsection{De Sitter Domain Walls}\label{dsw}
Our simplest example of a CDL geometry describes a one-parameter family of (thick) domain walls interpolating between two degenerate dS minima.  Geometries that describe genuine decays are given later, as they are in general more complicated. The domain walls are simple because in these cases $a(\xi)$ is symmetric under $\xi\leftrightarrow\pi-\xi$ and therefore only involves $\sin\xi$ when written in the form \eqref{sc}.  Our domain walls are given by
\begin{equation}\label{dswx}
a(\xi)=c\sqrt{1-\sqrt{1-\frac{2}{c^2}\sin^2\xi}},
\end{equation}
where $c$ is any constant greater than $\sqrt2$, as required by the reality of the inner square root. As $\xi$ approaches $0$ or $\pi$ we may expand the inner square root and check the smoothness conditions \eqref{bcat0} and \eqref{bcat1}. We also need to check the null energy condition \eqref{enec}, which reads
\begin{equation}\label{dswxn}
a'^2-aa''-1=\frac{(c^2-2)(c-\sqrt{c^2-2\sin^2\xi})}{(c^2-2\sin^2\xi)^{3/2}}\sin^2\xi\ge0,
\end{equation}
which is manifestly satisfied for any $c\ge\sqrt2$.

Next, we analytically continue \eqref{dswx} to Lorentzian signature by applying \eqref{con1}. The resulting scale factor is
\begin{equation}\label{dswt}
\hat a_1(t)=c\sqrt{\sqrt{1+\frac{2}{c^2}\sinh^2t}-1},
\end{equation}
whose late time behavior is
\begin{equation}\label{dswta}
\hat a_1(t)\to\sqrt{\frac{c}{\sqrt2}}e^{t/2}\quad
\text{as}\quad t\to\infty.
\end{equation}
Matching this to \eqref{ldsa}, we find an asymptotically dS space with radius $\ell_1=2$. This agrees with the bound $\ell_2>1$ mentioned earlier and proven in Appendix \ref{dsrad} (here $\ell_1=\ell_2$). We still need to check the null energy condition \eqref{lnec}, which is simply the analytic continuation of \eqref{dswxn}:
\begin{equation}\label{dswtn}
\dot{\hat a}_1^2-\hat a_1\ddot{\hat a}_1-1=\frac{(c^2-2)(\sqrt{c^2+2\sinh^2t}-c)}{(c^2+2\sinh^2t)^{3/2}}\sinh^2t\ge0.
\end{equation}
By symmetry the second analytic continuation \eqref{con2} gives exactly the same $\hat a_2(t)=\hat a_1(t)$.

We have verified that \eqref{dswx} satisfies all conditions (a)--(g) to be a CDL geometry. We may then use \eqref{phiVeq} to solve for $\phi(\xi)$ and $V(\phi)$. Their analytic forms are not very illuminating, as $\phi(\xi)$ involves a hypergeometric function and $V(\phi)$ is written in terms of the inverse of $\phi(\xi)$. However, in many applications of CDL (such as calculating the correlation functions) it is $a(\xi)$ that we would like to be simple, not $V(\phi)$.


\subsection{Decays from dS to dS}
In Sec.~\ref{dsw} we discussed domain walls interpolating between two degenerate dS minima. When this degeneracy is broken, we arrive at the phenomenologically interesting case \cite{Freivogel:2005vv} of decays from one dS space to another with a smaller cosmological constant. We give analytic examples of such geometries in this subsection.

In order to interpolate between two dS minima with different cosmological constants, it is necessary to break the symmetry of $a(\xi)$ under $\xi\leftrightarrow\pi-\xi$. In the form \eqref{sc} this means that $a(\xi)$ has to depend on $\cos\xi$ in addition to $\sin\xi$. Under the analytic continuations \eqref{con1} and \eqref{con2}, $\cos\xi$ becomes $\cosh t$ and $-\cosh t$ respectively as we see in \eqref{shch}. This minus sign is crucial in making the late time behaviors of $\hat a_1(t)$ and $\hat a_2(t)$ different, which is necessary for them to describe asymptotically dS spaces with different cosmological constants.

\begin{figure}[!htbp]
\includegraphics[width=\textwidth]{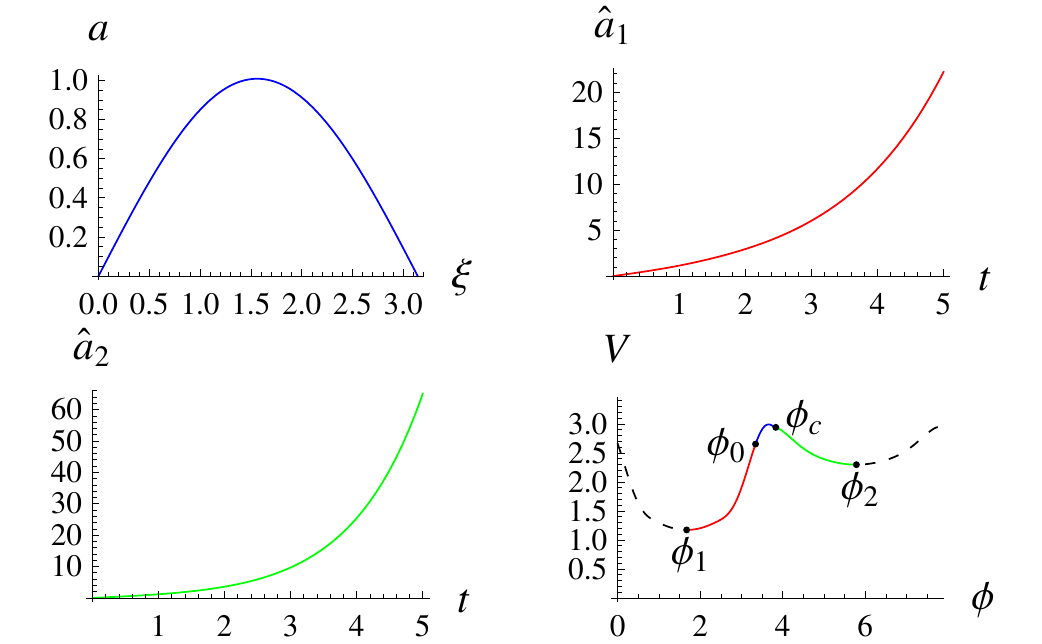}
\caption{A CDL geometry describing the decay from dS to dS. We have the Euclidean geometry $a(\xi)$ on the top left, the first FRW geometry $\hat a_1(t)$ that asymptotes to the true dS vacuum on the top right, the second FRW geometry $\hat a_2(t)$ describing the parent dS on the bottom left, and the scalar potential $V(\phi)$ on the bottom right. The blue, red, and green segments are traversed by the scalar field in the three regions of the CDL geometry as depicted in Figure~\ref{fig:continuation}. The black dashed lines are ``conjectures'' from what we expect qualitatively -- the potential there cannot be solved numerically from the CDL geometry because the field never goes there. If one could solve the potential analytically between the two minima, it can then be continued to the dashed region.}
\label{fig:ds2ds}
\end{figure}

One of the simplest examples that we found is
\begin{equation}\label{dsx}
a(\xi)=\left(\frac{1+10^{1/4}}{1+\left[10-\sin^2\xi\left(\sqrt{2-\sin^2\xi}+\cos\xi\right)\right]^{1/4}}\right)^{1/2}\sin\xi.
\end{equation}
One can show that this is of the form \eqref{sc} by first writing out its Taylor expansion in terms of $\sin\xi$ and $\cos\xi$, and then eliminating all quadratic or higher order terms in $\cos\xi$ by applying $\cos^2\xi=1-\sin^2\xi$. One can check the null energy conditions \eqref{enec} and \eqref{lnec}. Upon analytic continuation we have
\begin{equation}
\hat a_{1,2}(t)=\left(\frac{1+10^{1/4}}{1+\left[10+\sinh^2t\left(\sqrt{2+\sinh^2t}\pm\cosh t\right)\right]^{1/4}}\right)^{1/2}\sinh t,
\end{equation}
where $\hat a_1(t)$ takes the plus sign and asymptotes to
\begin{equation}
\hat a_1(t)\to\frac{\sqrt{1+10^{1/4}}}{2^{3/4}}e^{5t/8}\quad
\text{as}\quad t\to\infty,
\end{equation}
and $\hat a_2(t)$ takes the minus sign and asymptotes to
\begin{equation}
\hat a_2(t)\to\frac{\sqrt{1+10^{1/4}}}{2^{3/4}}e^{7t/8}\quad
\text{as}\quad t\to\infty.
\end{equation}
The radii of the two dS vacua are $\ell_1=8/5$ and $\ell_2=8/7$, which again obey the bound of Appendix \ref{dsrad}. We show this geometry in $d=4$ in Figure~\ref{fig:ds2ds}, together with the scalar potential that gives rise to it.

A family of geometries of this type can be found by varying the parameters in \eqref{dsx} (subject to the smoothness and null energy conditions). We will not give the exact parameter space here, but we note that it is quite large. For instance, one can verify that all constraints are still satisfied if we change both constants ``10'' in \eqref{dsx} to any number larger than $4.2$, or if we change the power $1/4$ to any number between 0 and $1/4$.

\subsection{Decays from dS to Minkowski Space}
\begin{figure}[!htbp]
\includegraphics[width=\textwidth]{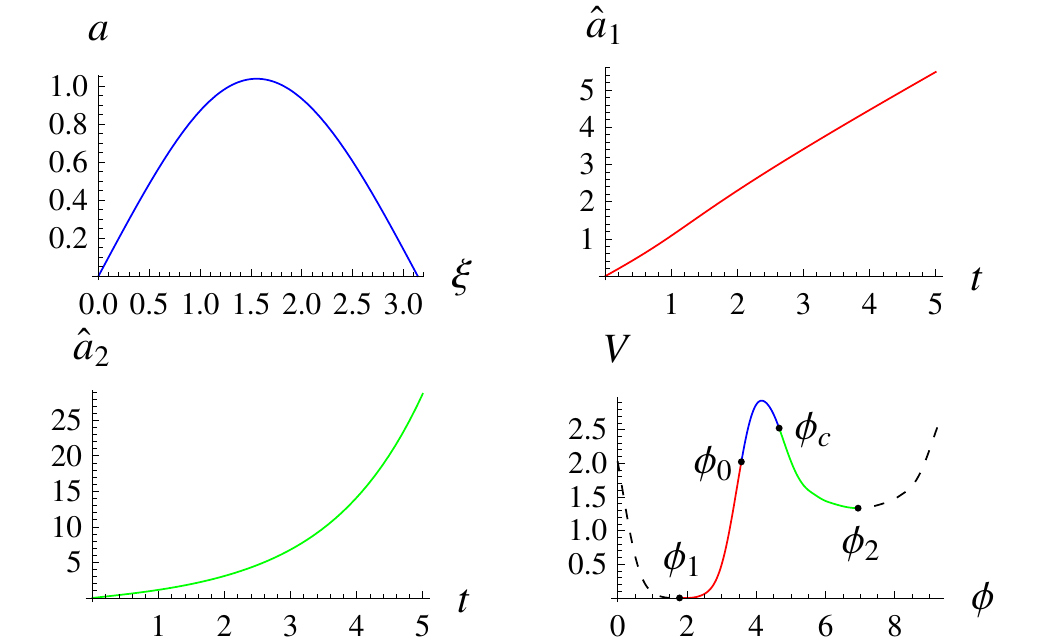}
\caption{The geometry and potential describing a decay from dS to asymptotically Minkowski space. See the caption of Figure~\ref{fig:ds2ds} for detailed explanations.}
\label{fig:ds2flat}
\end{figure}

In this subsection we consider decays from dS to asymptotically Minkowski space.  These are potentially interesting for conceptual reasons, as explained in \cite{Susskind:2007pv,crunchhat,Bousso:2011up}.  A ``simple'' geometry of this type is
\begin{equation}\label{frwx}
a(\xi)=\frac{\frac32\sin\xi}{1+\left[8-\sin^2\xi\left(\sqrt{2-\sin^2\xi}+\cos\xi\right)\right]^{1/3}}+\arcsin\left(\frac{\sin\xi}{2}\right).
\end{equation}
One can check that it satisfies the smoothness conditions \eqref{bcat0}, \eqref{bcat1} and the null energy conditions \eqref{enec}, \eqref{lnec}. Upon analytic continuation we have
\begin{equation}
\hat a_{1,2}(t)=\frac{\frac32\sinh t}{1+\left[8+\sinh^2t\left(\sqrt{2+\sinh^2t}\pm\cosh t\right)\right]^{1/3}}+\text{arcsinh}\left(\frac{\sinh t}{2}\right),
\end{equation}
where $\hat a_1(t)$ takes the plus sign and the exponentially growing terms cancel, leading to
\begin{equation}
\hat a_1(t)\to t\quad
\text{as}\quad t\to\infty,
\end{equation}
and $\hat a_2(t)$ takes the minus sign and asymptotes to
\begin{equation}
\hat a_2(t)\to\frac{3}{2^{4/3}}e^{2t/3}\quad
\text{as}\quad t\to\infty.
\end{equation}
The radius of the parent dS space is therefore $\ell_2=3/2$, again satisfying the bound of Appendix \ref{dsrad}. We show this geometry and its potential in $d=4$ in Figure~\ref{fig:ds2flat}.

As before, a family of geometries of this type can be found by varying the parameters in \eqref{frwx}. Additionally, we could get CDL geometries that describe decays from dS to FRW which has a zero cosmological constant but does not have the asymptotic behavior \eqref{lflata}. As discussed above \eqref{lflata} this means that the scalar field is rolling off to infinity in the FRW. A particular class of interesting FRW solutions of this type was studied in \cite{Dong:2011uf} and conjectured to have holographic duals. They are characterized by the following late time behavior:
\begin{equation}\label{lcfrw}
\hat a_1(t)\to ct\quad
\text{as}\quad t\to\infty.
\end{equation}
We give analytic CDL geometries of this type by multiplying the ``$\arcsin$'' term in \eqref{frwx} by a constant $c>1$, and multiplying the first term by $(2-c)$ to preserve the smoothness conditions.\footnote{The lower limit $c>1$ is required by the null energy condition \eqref{lnec}. The apparent upper limit $c<2$ is superficial and can be relaxed by changing the ``2'' inside the $\arcsin$ in \eqref{frwx}.} Over a range of $c$ the null energy conditions are satisfied, and we get an asymptotically linear scale factor $\hat a_1(t)\to ct$ with $c>1$.

\subsection{Noncompact Examples}
\begin{figure}[!htbp]
\includegraphics[width=\textwidth]{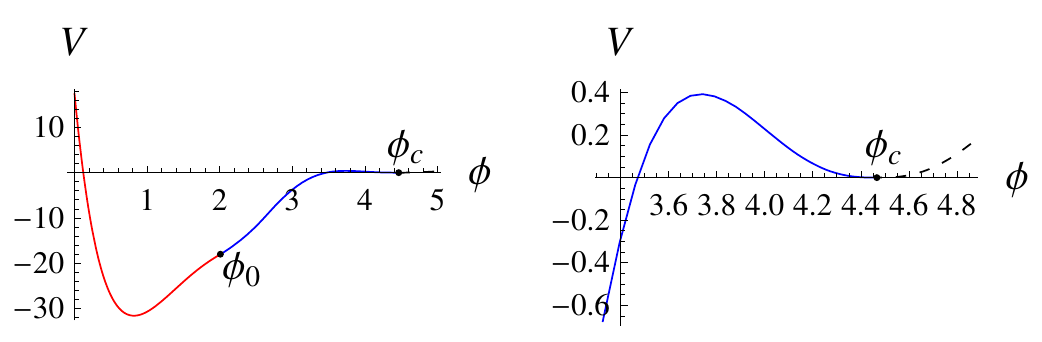}
\caption{The potential $V(\phi)$ leading to a CDL geometry \eqref{flat2ads4d} describing the decay from Minkowski space to a crunching AdS. On the right we zoom in on the same potential at $\phi_c$ (the asymptotic field value) and see that it is a local minimum with a very shallow barrier.}
\label{fig:flat2ads}
\end{figure}

Noncompact examples have simpler functional forms, but have the added complication of the conditions \eqref{ncVmin} or \eqref{bfbound}.  These decays are arguably the least interesting, since exactly Minkowski spaces are expected to be supersymmetric and stable, and metastable AdS is inherently ill-defined \cite{Horowitz:2007pr,metads,brads}.  A candidate example for a decay of Minkowski space to a crunch is
\be
a(\xi)=\xi\left(1+\frac{\xi^2}{(1+\xi^2)^{3/2}}\right).
\ee
The radical is still necessary because of the argument given below equation \eqref{deltabound}.  This obeys the null energy conditions \eqref{enec} and \eqref{lnec}, and crunches inside the bubble at finite time.  For $d=4$ however it corresponds to rolling down from a maximum, and we find we need to set $d=3$ to satisfy \eqref{ncVmin}.  This can be checked analytically by expanding \eqref{ncVmin} to order  $1/\xi^6$ at large $\xi$.\footnote{Note that for the compact examples in the previous subsections the dimension was irrelevant.  It is only for noncompact geometries that we have the inequalities \eqref{ncVmin} or \eqref{bfbound} which depend on dimension.}  An improved example that works in $d=4$ is
\be\label{flat2ads4d}
a(\xi)=\xi\left(1+\frac{\xi^2}{\sqrt{1+\xi^6}}\right).
\ee
We can show that the potential leading to this geometry has a metastable minimum by expanding \eqref{ncVmin} to order $1/\xi^{10}$.  This potential is shown in Figure \ref{fig:flat2ads}.

\begin{figure}[!htbp]
\includegraphics[width=\textwidth]{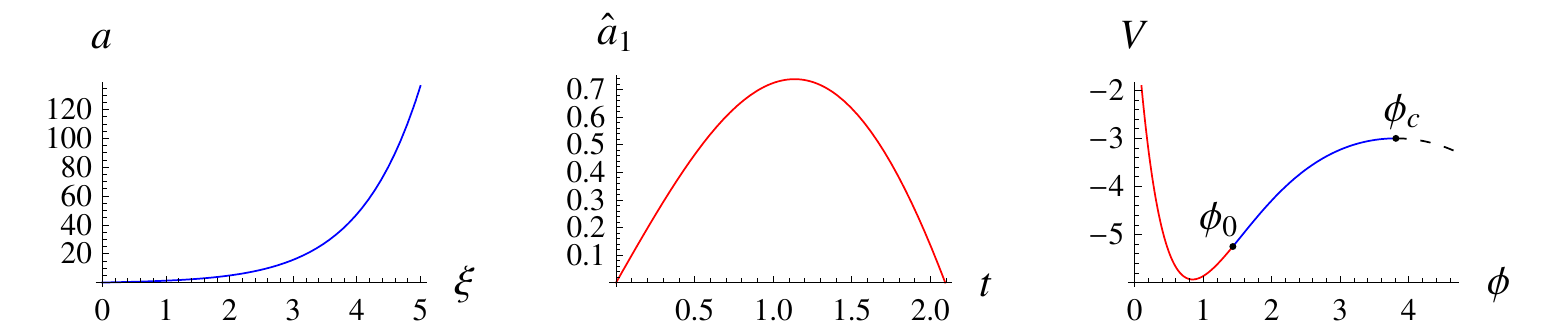}
\caption{The geometry and potential for a decay from AdS to a crunch.}
\label{fig:ads2ads}
\end{figure}

A candidate family of decays of AdS to a crunch is
\be
a(\xi)=(1+c)\sinh \xi-2c\sinh\frac{\xi}{2}.
\ee
This example is simple enough that we can check the null energy conditions \eqref{enec} and \eqref{lnec} analytically, finding that any $c>0$ is allowed.  It is also not hard to check that the BF bound (\ref{bfbound}) is satisfied for $d\ge3$.  This geometry and its potential are shown in Figure \ref{fig:ads2ads} for $c=1$, $d=4$.

\section*{Acknowledgments}
We thank A. Brown, B. Horn, M. Salem, E. Silverstein, D. Stanford, L. Susskind, and G. Torroba for helpful discussions.  XD would also like to thank the Kavli Institute for Theoretical Physics for hospitality during the final phase of this work.  This research was supported in part by the NSF under grant PHY-0756174.  XD is also supported by the DOE under contract DE-AC03-76SF00515. DH is also supported by a Stanford Graduate Fellowship and the Howrey Term Endowment Fund in Memory of Dr. Ronald Kantor.

\appendix
\section{Null Energy Condition}
\label{necapp}
In this appendix we show the equivalence of the inequalities (\ref{enec}) and (\ref{lnec}) to the null energy condition
\be
\label{nec}
T_{\mu\nu}k^\mu k^\nu\geq0,
\ee
for any null $k^{\mu}$.  We first apply this to the ``warped de Sitter'' geometry (\ref{warpedds}), which we rewrite as
\be
ds^2=d\xi^2+a(\xi)^2 \gamma_{ij}dx^i dx^j.
\ee  
The Ricci tensor is
\begin{align}\nonumber
R_{\xi\xi}&=-(d-1)\frac{a''}{a}\\\nonumber
R_{ij}&=-\left[a a''+(d-2)(a'^2-1)\right]\gamma_{ij}\\
R_{\xi i}&=R_{i \xi}=0,
\end{align}
so the Einstein tensor is
\begin{align}\nonumber
G_{\xi\xi}&=\frac{(d-1)(d-2)}{2}\left[\left(\frac{a'}{a}\right)^2-\frac{1}{a^2}\right]\\\nonumber
G_{ij}&=\left[(d-2)\frac{a''}{a}+\frac{(d-2)(d-3)}{2}\left(\left(\frac{a'}{a}\right)^2-\frac{1}{a^2}\right)\right]a^2 \gamma_{ij}\\
G_{\xi i}&=G_{i \xi}=0.
\end{align}
To check the null energy condition, consider the radial null vector $k=\partial_\xi+\frac{1}{a}\partial_\omega$.\footnote{Any nonradial vector can be boosted into a radial one unless it lies entirely within the dS slice, but in that case the Null Energy Condition is trivially saturated since the dS Einstein tensor is proportional to the metric.}  Contracting this with $G_{\mu\nu}$ we find (\ref{nec}) is satisfied if
\be
\left(\frac{a'}{a}\right)^2-\frac{1}{a^2}-\frac{a''}{a}\geq0,
\ee
which agrees with (\ref{enec}).  We can easily continue this Einstein tensor to find the Einstein tensor for the open FRW metric (\ref{frw}), and a similar computation confirms (\ref{lnec}).

\section{A Bound on Parent dS Radius}\label{dsrad}
In this appendix we show that for any compact CDL geometry describing the decay of dS space, the parent dS radius (which is $\ell_2$ according to \eqref{ldsa2}) must be greater than $\xi_c/\pi$. In the convenient units proposed in Sec.~\ref{comp} where $\xi_c=\pi$, this means $\ell_2>1$. We will not need to know what kind of space the parent dS decays into, whether it is dS, AdS, Minkowski space, or even something else.

We prove this by using the null energy conditions \eqref{enec} and \eqref{lnec}. First, we note that it is impossible for $a(\xi)$ or $\hat a(t)$ to have a local minimum within their physical domains. Clearly if for example $a(\xi)$ had a local minimum, we would have $a'(\xi)=0$ and $a''(\xi)\ge0$ there, manifestly violating the null energy condition \eqref{enec}. Therefore it is impossible to have a contracting phase followed by an expanding phase. Combining this with the boundary conditions \eqref{bcat0} and \eqref{bcat1}, we find that $a(\xi)$ must monotonically increase to a maximum at some $\xi_m$ and after that monotonically decrease to 0. For $\hat a_2(t)$ there has to an expanding phase \eqref{ldsa2} at late times, so $\hat a_2(t)$ must be a monotonically increasing function for all $t\ge0$.

Let us define an ``energy function'' $E\equiv(a'^2-1)/a^2$ and take its derivative
\begin{equation}
\frac{dE}{d\xi}=\frac{d}{d\xi}\left(\frac{a'^2-1}{a^2}\right)=-\frac{2a'(a'^2-aa''-1)}{a^3}.
\end{equation}
Using the null energy condition \eqref{enec}, we find that $E$ never increases in an expanding phase and never decreases in a contracting phase. For a more physical argument, we note that $E$ is equal to the total energy up to a constant coefficient, as we can see from the second equation in \eqref{eom}.\footnote{The Euclidean equations of motion \eqref{eom} are completely the same as the Lorentzian equations \eqref{leom} with the inverted potential $-V(\phi)$, so we may use our usual intuition about real-time evolution.} The total energy is drained by friction during an expanding phase and replenished  by anti-friction during a contracting phase. Exactly the same statements hold for $\hat E_2\equiv(\dot{\hat a}_2^2-1)/\hat a_2^2$.

It is therefore clear that $E(\xi)$ reaches an absolute minimum exactly as $a(\xi)$ reaches its maximum at $\xi_m$:
\begin{equation}\label{ebound}
E(\xi)\ge E(\xi_m)=-\frac{1}{a(\xi_m)^2}\qquad
\text{for}\qquad\forall\xi\in[0,\xi_c].
\end{equation}
Let us abbreviate $a(\xi_m)$ as $a_m$. For the expanding phase $0\le\xi\le\xi_m$ we rewrite the above inequality as
\begin{equation}\label{aprime}
\frac{a'}{\sqrt{1-a^2/a_m^2}}\ge1
\end{equation}
and integrate both sides from $\xi=0$ to $\xi=\xi_m$, which gives
\begin{equation}\label{exp}
\frac{\pi}{2}a_m\ge\xi_m.
\end{equation}
For the contracting phase $\xi_m\le\xi\le\xi_c$ there is a minus sign on the left hand side of \eqref{aprime}. After integrating from $\xi_m$ to $\xi_c$ we find
\begin{equation}
\frac{\pi}{2}a_m\ge\xi_c-\xi_m.
\end{equation}
Combining this with \eqref{exp} we arrive at
\begin{equation}\label{xic}
\pi a_m\ge\xi_c.
\end{equation}

Finally, we argue that $a_m$ is bounded from above by $\ell_2$. From the analytic continuation \eqref{con2} we find
\begin{equation}
E|_{\xi=\xi_c}=\left.\frac{a'^2-1}{a^2}\right|_{\xi=\xi_c}=-\left.\frac{\dot{\hat a}_2^2-1}{\hat a_2^2}\right|_{t=0}=-\hat E_2|_{t=0}.
\end{equation}
Physically, the field is at rest when $\xi=\xi_c$ or $t=0$, so the total ``Euclidean energy'' $-V(\phi_c)$ is just minus the ``Lorentzian energy''. From this we have a chain of (in)equalities
\begin{equation}
-\frac{1}{a_m^2}=E|_{\xi=\xi_m}\le E|_{\xi=\xi_c}=-\hat E_2|_{t=0}\le-\hat E_2|_{t\to\infty}=-\frac{1}{\ell_2^2},
\end{equation}
where we have the first inequality because $E$ never decreases in a contracting phase, and the second inequality because $\hat E_2$ never increases in an expanding phase. The last equality comes from the asymptotically dS condition \eqref{ldsa2}. Therefore we have
\begin{equation}\label{ambound}
a_m\le\ell_2.
\end{equation}
Combining this with \eqref{xic}, we have shown the bound on the parent dS radius
\begin{equation}
\ell_2>\frac{\xi_c}{\pi}.
\end{equation}
This inequality cannot be saturated without saturating all the above inequalities. In particular, saturating \eqref{ebound} and \eqref{ambound} leads to $a(\xi)=\ell_2\sin(\xi/\ell_2)$ which describes a pure dS geometry rather than a CDL decay.

\bibliographystyle{jhep}
\bibliography{bibliography}
\end{document}